\documentclass[aps,prl,showpacs,twocolumn]{revtex4}
\usepackage{multirow}
\usepackage{graphicx}
\usepackage{amsmath}
\usepackage{pslatex}

\newcommand{\sqrtsNN}{\mbox{$\sqrt{\mathrm{s}_{_{\mathrm{NN}}}}$}}

\newcommand{\pt}{$p_{\rm T}$}
\newcommand{\DDbar}{${\rm D}\overline{\rm D}$}
\def \auau  {Au+Au}

\begin{document}
\title{\DDbar\ Correlations as a Sensitive Probe for Thermalization
in High Energy Nuclear Collisions}
\date{\today}

\author{
X.~Zhu$^{1,2,3}$,
M.~Bleicher$^{1}$,
S.L.~Huang$^{4}$,
K.~Schweda$^{5}$,
H.~St\"ocker$^{1,2}$,
N.~Xu$^{4}$,
P.~Zhuang$^{3}$
}

\affiliation{$^1$ Institut f\"ur Theoretische Physik, Johann Wolfgang Goethe-Universit\"at, Max-von-Laue-Str.~1, D-60438 Frankfurt am Main, Germany} 
\affiliation{$^2$ Frankfurt Institute for Advanced Studies (FIAS), Max-von-Laue-Str.~1, D-60438 Frankfurt am Main, Germany}
\affiliation{$^3$ Physics Department, Tsinghua University, Beijing 100084, China}
\affiliation{$^4$ Nuclear Science Division, Lawrence Berkeley National Laboratory, Berkeley, CA 94720, USA}
\affiliation{$^5$ Physikalisches Institut, Universit\"at Heidelberg, Philosophenweg 12, D-69120 Heidelberg, Germany}

\begin{abstract}
We propose to measure azimuthal correlations of heavy-flavor hadrons to address
the status of thermalization at the partonic stage of light quarks and
gluons in high-energy nuclear collisions.
In particular, we show that hadronic interactions at the late stage
cannot significantly disturb the initial back-to-back azimuthal correlations of \DDbar\
pairs. Thus, a decrease or the complete absence of these initial
correlations does indicate frequent interactions of heavy-flavor
quarks and also light partons in the partonic stage, which are essential
for the early thermalization of light partons.
\end{abstract}

\pacs{25.75.-q}

\maketitle

{\it Introduction}

Lattice QCD calculations, at vanishing or finite net-baryon density,
predict a cross-over transition from the deconfined thermalized partonic
matter (the Quark Gluon Plasma, QGP) to hadronic matter at a critical
temperature $T_{\rm c} \approx 150$--180~MeV~\cite{karsch}.

Measurements of hadron yields in the intermediate and high transverse
momentum (\pt) region indicate that dense matter is produced in
\auau\ collisions at RHIC~\cite{star_white,phenix_white}. The
experimentally observed large amount of elliptic flow of multi-strange
hadrons~\cite{star_omgeav2}, such as the $\phi$ meson and the $\Omega$
baryon, suggest that collective motion develops in
the early partonic stage of the matter produced in these collisions. A
crucial issue to be addressed next 
is the thermalization status of this partonic matter.

\begin{figure} [h]
\includegraphics[width=0.43\textwidth]{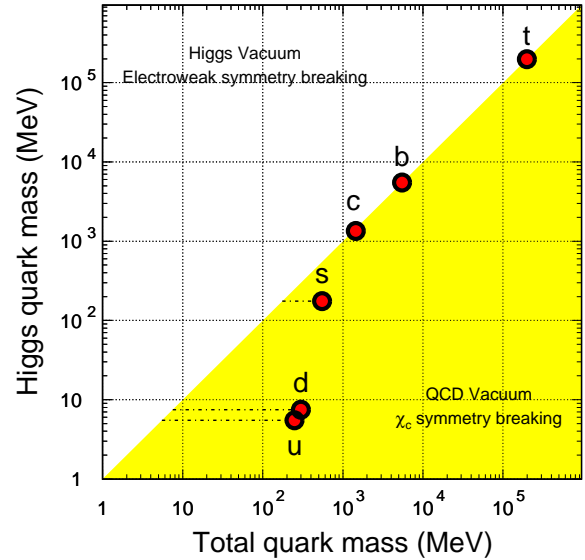}
\caption{(Color online) Quark masses in the QCD vacuum and the Higgs vacuum. A
  large fraction of the light quark masses is due to the chiral
  symmetry breaking in the QCD vacuum. The numerical values
  were taken from Ref.~\cite{pdg:2004}.}. \label{fig1}
\end{figure}

Heavy-flavor (c, b) quarks are particularly excellent tools~\cite{Svetitsky:1996nj,
Svetitsky:1997xp,Svetitsky:1997bt} to study the
thermalization of the initially created matter. As shown in
Fig.~\ref{fig1}, their large masses are almost exclusively generated
through their coupling to the Higgs field in the electro-weak sector,
while masses of light quarks (u, d, s) are dominated by spontaneous
breaking of chiral symmetry in QCD. This means that in a QGP, where
chiral symmetry might be restored, light quarks are left with their bare
current masses while heavy-flavor quarks remain heavy. Due to their
large masses ($\gg\Lambda_{QCD}$), the heavy quarks can only be
pair-created in early stage pQCD processes. Furthermore, their production cross
sections in nuclear collisions are found to scale with the number of
binary nucleon-nucleon collisions~\cite{Adler:2004ta,starcharm_new}. In the subsequent
evolution of the medium, the number of heavy quarks is conserved because the
typical temperature of the medium is much smaller than the heavy
quark (c, b) masses, resulting in negligible secondary pair production. In addition,
the heavy quarks live much longer than the lifetime of the
formed high-density medium, decaying well outside. These
heavy quarks (c, b) can participate in collective motion or even
kinetically equilibrate if, and only if, interactions at the partonic level
occur at high frequency. The idea of statistical hadronization of
kinetically equilibrated charm quarks~\cite{pbm_3} even predicted
significant changes in hidden charm hadron
production~\cite{pbm_charm}.  
Hence, heavy-flavor quarks are an ideal probe to
study early dynamics in high-energy nuclear collisions.

Recent STAR and PHENIX results on elliptic flow and nuclear
modification factors of non-photonic electrons indicate that charm
quarks might indeed participate in the collective motion of the matter
produced in Au+Au collisions at RHIC.
To explain the data, a
large drag diffusion coefficient in the Langevin equation, describing
the propagation of charm quarks in the medium, is found to be
necessary \cite{Moore:2004tg}.  Two- and three-body
interactions~\cite{wicks_05,liu_06} of heavy-quarks and resonant
rescattering~\cite{hees_05} in the partonic stage seem to become important.
These investigations suggest that heavy-quarks actively participate in
the partonic stage.

In this paper, we study the change of azimuthal correlations of 
D and $\overline{\rm D}$ meson pairs as a sensitive indicator of
frequent occurrences of partonic scattering. Since heavy-flavor quarks
are pair-created by initial scattering processes, such as $gg\rightarrow
{\rm c}\overline{\rm c}$, each quark-antiquark pair is correlated in
relative azimuth $\Delta \phi$ due to momentum conservation. In
elementary collisions, these correlations survive the fragmentation
process to a large extent and hence are observable in the 
distribution of the relative azimuth of
pairs of D and $\overline{\rm D}$ mesons.  We
evaluate by how much these correlations should be
affected by the early QGP stage and by the hadronic scattering processes
in the late hadronic stage.

{\it Results and Discussions}

\begin{figure} [h]
\includegraphics[width=0.45\textwidth]{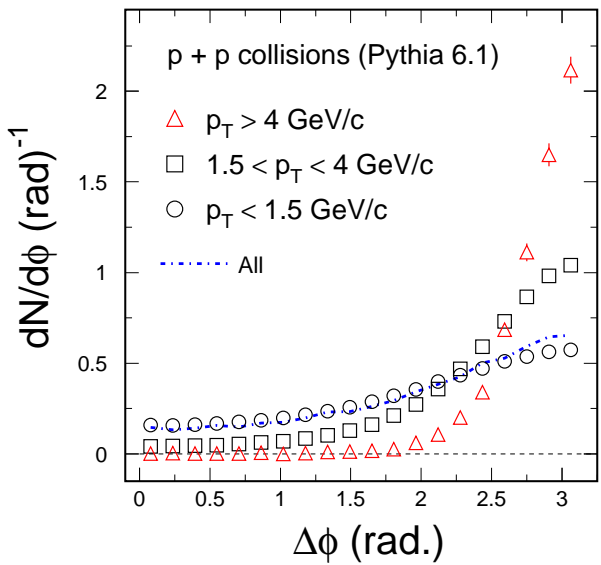}
\caption{(Color online) Correlations in relative azimuth, $\Delta
  \phi$, of ${\rm D}\overline{\rm D}$ pairs from pp collisions at
  $\sqrt{s} = 200$~GeV, as calculated by PYTHIA (v.~6.139) for different
  ranges of \emph{single} D meson \pt.} \label{fig2}
\end{figure}

We start by reviewing the D and $\overline{\rm D}$ angular
correlations in pp collisions.  The Monte Carlo event generator 
PYTHIA~\cite{Sjostrand:2000wi}
reproduces
well the experimentally observed correlations of D mesons, measured at
fixed target energies~\cite{hermine}.  Figure~\ref{fig2} shows 
such correlations as calculated by PYTHIA (v.~6.139)
for pp collisions at $\sqrt{s} = 200$~GeV. The $\Delta \phi$
distribution is peaked at 180$^\circ$, especially for high \pt\ D
mesons, as one would expect for back-to-back pairs stemming from hard 
scatterings of partons.

To explore how the QCD medium generated in central ultra-relativistic
nucleus-nucleus collisions influences the correlations of D and
$\overline{\rm D}$ meson pairs, we employ a non-relativistic Langevin approach
which describes the random walk of charm quarks in a QGP and was first
described in
Refs.~\cite{Svetitsky:1996nj,Svetitsky:1997xp,Svetitsky:1997bt},
\begin{equation}
\frac{d\vec{p}}{dt} =-\gamma(T)\vec{p}+\vec{\eta},
\end{equation}
where $\vec{\eta}$ is a Gaussian noise variable, normalized such that
$\langle \eta_i(t)\eta_j(t')
\rangle=\alpha(T)\delta_{ij}\delta(t-t')$ with $i,j$ indexing directions. Both the drag coefficient
$\gamma$ and the momentum-space diffusion coefficient $\alpha$ depend
on the \emph{local} temperature, $T$. We use the same parameterization for
$\gamma$ as in Ref.~\cite{Svetitsky:1996nj}, $\gamma(T) = a T^2$, with
$a = 2\cdot 10^{-6}$ (fm/$c$)$^{-1}$ MeV$^{-2}$ and also neglect a
momentum dependence of $\gamma$. $\gamma$ and $\alpha$ are related by
the fluctuation-dissipation relation in equilibrium, $\langle p_i^2
\rangle=\alpha/2\gamma$. As in Ref.~\cite{Svetitsky:1996nj},
$\alpha$ is calculated from the above equation with $\langle
p_i^2\rangle=1.33\, m_{\rm c}\, T$, with a charm quark mass $m_{\rm
c}=1.5$~GeV/$c^2$.

For simplicity, the initial conditions for central collisions of
heavy nuclei (Au or Pb) are the same as in
Ref.~\cite{Svetitsky:1996nj}. We assume a plasma in thermal
equilibrium occupying a cylinder of fixed radius $R$, equal to the
radius of the colliding nuclei ($R=7$~fm in the present
calculations), at the time of its full formation $\tau_0$. After
that, the plasma evolves according to Bjorken's
hydrodynamics, with a temperature dropping like $T=T_0\,(\tau_0/\tau)^{1/3}$.
Before the time $\tau_0$, the plasma is changing rapidly and is not
fully equilibrated. We assume (as done in
Ref.~\cite{Svetitsky:1996nj})
that the plasma is already in equilibrium before $\tau_0$, with
$T=T_0$ and with the
charmed quarks diffusing in the same way as they do after $\tau_0$.
In order to isolate the effects purely due to parton-parton re-scattering in the
outlined medium, we generated c and $\overline{\rm c}$ quarks with the same $p_{\rm T}$ and zero longitudinal momentum back-to-back, i.e.\ with
$\Delta \phi = \pi$ at a time of the order of
$1/m_c\simeq$~0.1~fm/$c$, and used a delta function for fragmenting the
charm quark into a charmed hadron at the hadronization stage.  The
radius $r$ where each pair is created is randomly generated from a distribution
reflecting the number of binary nucleon-nucleon collisions taking
place at that radius, $p(r){\rm d}\,r\propto(R^2-r^2)\,2\pi\, r\, {\rm d}r$.
The angular distribution of the initially produced charm quarks
in the transverse plane is isotropic.

\begin{figure} [h]
\includegraphics[width=0.45\textwidth]{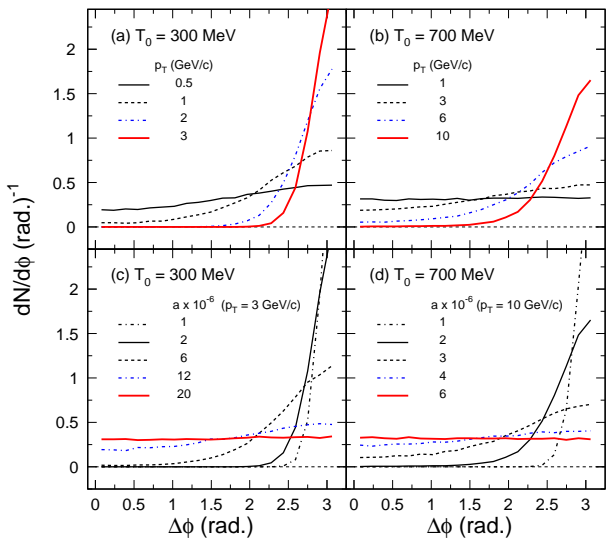}
\caption{(Color online) Correlations in relative azimuth $\Delta \phi$
  of \DDbar\ pairs from Langevin calculations with
  $T_0=300$~MeV (RHIC) and $700$~MeV (LHC).  The upper part shows the
  dependence of the correlations on the initial $p_{\rm T}$ of the c quark; 
  the lower part shows the drag
  coefficient dependence.} \label{fig3}
\end{figure}

The evolution of the charm momenta with the Langevin equation is
stopped when $T$ reaches the critical temperature $T_{\rm c}=165$~MeV
or when the charm quark leaves the QGP volume. Furthermore, to get a first estimation
of the QGP effect on the charm quark pairs azimuthal correlation, we
omit the possible contribution of the mixed
phase. Figure~\ref{fig3}\,(a) shows the results for the D meson (charm
quark) pairs angular correlations for different initial charm quark $p_{\rm
T}$ values, for $T_0=300$~MeV and
$\tau_0=0.5$~fm/$c$ (typical values for RHIC collisions). 
We see that the fastest charm quarks (represented by the $p_{\rm
T}=3$~GeV/$c$ red curve) are able to escape from the QGP without suffering
significant medium effects, while the slower quarks (see the $p_{\rm
T}=0.5$~GeV/$c$ black line) have their pair azimuthal correlation almost
completely smeared out by the interactions in the medium.

Figure~\ref{fig3}\,(b) shows the corresponding results for $T_0=700$~MeV and
$\tau_0=0.2$~fm/$c$, values representative of LHC energies. Here, the interactions of
the charm quarks with the medium are so frequent that only the most
energetic charm quarks preserve part of their initial angular
correlation; low $p_{\rm T}$ pairs can even be completely stopped by
the medium. Because the c and the $\overline{\rm c}$ quarks of a given
pair are created together, in the
same space point, the pair has a higher escape probability
if both quarks escape the medium from the side of 
the fireball where the 
thickness is smaller. Thus, the $\Delta \phi$ distribution is shifted towards
smaller values. Presently, there is only longitudinal flow in the
hydro-dynamical model used in the calculations. We expect that the
strong radial flow will further enhance the same side correlations in the
case of fully kinetically thermalized charm pairs. Because the two
quarks of a pair
are created in the same position in the medium, they will be pushed in the
same direction by the radial flow. In fact, this mechanism is a known 
non-flow effect in the
measurement of collective flow~\cite{Voloshin:2003ud}.

The above results were obtained with a drag coefficient estimated with
perturbative QCD adopting a large coupling constant $\alpha_s=0.6$~\cite{Svetitsky:1987gq}. 
The recent pQCD calculations~\cite{GolamMustafa:1997id,hees_05} show
a factor of 2-3 smaller drag coefficient. Besides, as shown in
Ref.~\cite{hees_05}, non-perturbative contributions that arise from
quasi-hadronic bound states in the QGP might be important. The
presence of these resonances at moderate QGP temperatures would result
in a much larger drag coefficient. Since exact values of the drag
coefficient from lattice QCD calculations do not exist, we show the
sensitivity of our calculations when varying the parameter $a$ within
reasonable values.  Figures~\ref{fig3}\,(c) and (d) show the drag
coefficient dependence of the c$\overline{\rm c}$ angular correlations
for high-energy c quarks: $p_{\rm T}=3$ and 10~GeV/$c$ for $T_0=300$ and
700~MeV, respectively.  The c$\overline{\rm c}$ angular correlations
are washed out when $a$ is increased by around a factor of 5 in the first
case and by around a factor of 2 in the second, with respect to the pQCD
value.

While the initial correlations of ${\rm c}\overline{\rm c}$ pairs are
clearly affected by the formed QCD medium, and sensitive to
its temperature and drag coefficient, it is important to evaluate to
which extent the loss of correlations can be mimicked by \emph{hadronic}
interactions.  Therefore, we have investigated by how much the
${\rm c}\overline{\rm c}$ correlations deteriorate due to
hadronic scattering in the late stage of high-energy nuclear
collisions. For this purpose, we used the microscopic hadronic
transport approach UrQMD v.~2.2~\cite{urqmd1,urqmd2}.

This microscopic transport approach is based on the covariant
propagation of constituent quarks and diquarks accompanied by mesonic
and baryonic degrees of freedom. It simulates multiple interactions of
in-going and newly produced particles, the excitation and
fragmentation of color strings, and the formation and decay of hadronic
resonances.  A phase transition to a QGP state is \emph{not} incorporated
into the model dynamics. In the latest version of the model,
PYTHIA (v.~6.139) was integrated to describe the energetic primary
elementary collisions~\cite{Bratkovskaya:2004kv}.  Measured particle
yields and spectra 
are well reproduced by the approach~\cite{Bratkovskaya:2004kv}.

In this model, D mesons stem from the very early stage high-energy
nucleon-nucleon collisions, calculated with PYTHIA. For their
propagation in the hadronic medium, we consider elastic scattering of
D mesons with all other hadrons. The hadronic scattering cross-section
for D mesons is generally considered to be small~\cite{Lin:2000jp} and we take 2~mb in
our calculation. To evaluate the sensitivity of the calculations to
this parameter, we have repeated them using a ten times higher 
cross section, 20~mb.
The final state angular distribution of the elastic scattering between
D mesons and other hadrons can affect
the ${\rm D}\overline{\rm D}$ correlations. 
An isotropic angular distribution in the scattering process, motivated
by the idea of intermediate resonances formation (essentially
D$+\pi\leftrightarrow$D$^*$), has the strongest effect on the charm
correlations. For comparison, we have also used a forward peaked
distribution, simulated according to the distribution $p(\cos\theta)\propto \exp(7\cos\theta)$, where $\theta$ is the scattering angle in the center of mass system of the
decaying resonance.

\begin{figure} [h]
\includegraphics[width=0.45\textwidth]{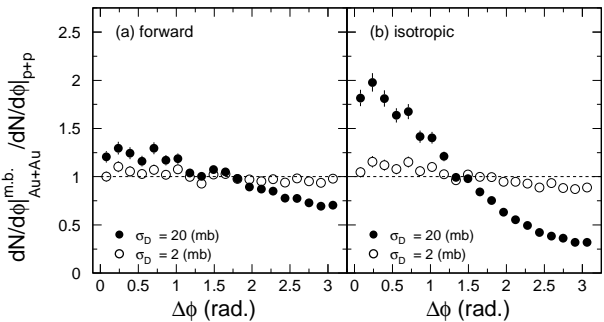}
\caption{Ratios between the $\Delta \phi$ distributions of \DDbar\
  pairs produced in minimum bias Au+Au reactions at \sqrtsNN$=200$~GeV, before and
  after hadronic rescattering, using forward (left panel) and
  isotropic (right panel) angular distributions.} \label{fig4}
\end{figure}

The results of the UrQMD calculations are shown in Fig.~\ref{fig4},
for minimum bias Au+Au reactions at \sqrtsNN$=200$~GeV. To better illustrate the
\emph{change} in the angular correlations of the ${\rm D}\overline{\rm
D}$ pairs, we show the ratio between the final distribution, affected
by the evolution with UrQMD, and the initial one, directly obtained
from PYTHIA (see Fig.~\ref{fig2}). For this specific analysis, all
${\rm D}\overline{\rm D}$ pairs were selected, irrespective of the
$p_{\rm T}$ of the D mesons. We observe that the ${\rm D}\overline{\rm
D}$ pair correlation, evaluated through the relative azimuthal angle
distribution, is barely affected by hadronic interactions. Except in
the somewhat drastic case of using a 20~mb hadronic cross-section and
even in this case only if we use an isotropic scattering angular
distribution (Fig.~\ref{fig4}\,(b)) the initial correlations are weakened.

We also find that most scatterings happen at the earliest times, when
the hadron density is very high. At RHIC energies, the (pure) hadronic
stage is presumably shorter than assumed in the present calculations.
Therefore, the hadronic stage should not induce visible changes to the
measured \DDbar\ angular correlations. Thus, a change of the ${\rm D}\overline{\rm D}$ angular correlation in heavy-ion
collisions, is dominated by frequent parton-parton
scatterings occurring in the QGP phase.

The calculations reported above show that heavy-flavor correlations
provide a sensitive tool to directly access the status of
equilibration of the partonic medium. A similar picture emerges from the
observation and suppression of back-to-back correlations of
unidentified high $p_{\rm T}$ hadrons (jets).  Both features originate
from the propagation of partons (heavy quarks or jets) in the medium.
However, D mesons (and B mesons) have the advantage that they can
still be identified, even if they lose a significant fraction of their
energy and get kinetically equilibrated. 

Compared to the corresponding balance functions~\cite{Bass:2000az} of
electric charge or strangeness, heavy-flavor correlations have 
obvious advantages: Heavy quarks and anti-quarks are pair created
in the early stage hard processes only and their subsequent annihilations are negligible. 
Hence, both numbers of heavy quarks and anti-quarks are conserved separately.
On the other hand, only the net-charge or total strangeness is conserved while the number of
positive (and negative) charges or strange and anti-strange particles can
be changed due to new pair production and annihilation throughout the
whole lifetime of the fireball.

{\it Conclusions and Outlook}

In summary, we argue that the observation of broadened angular
correlations of heavy-flavor hadron pairs in high-energy
heavy-ion collisions would be an indication of thermalization at the
partonic stage (among light quarks and gluons). We have seen that
hadronic interactions at a late stage in the collision evolution
cannot significantly disturb the azimuthal correlations of \DDbar\
pairs.  Thus, a visible decrease or the complete absence of such
correlations, would indicate frequent interactions of heavy-flavor
quarks and other light partons in the partonic stage, therefore imply early thermalization
of light quarks (which would be the main scattering centers) in nucleus-nucleus collisions at RHIC and LHC.

These measurements require good statistics of events in which
\emph{both} D mesons are cleanly reconstructed. A complete
reconstruction of the D mesons (i.e.\ of \emph{all} their decay
products) in full azimuth is essential to preserve the kinematic
information and to optimize the acceptance for detecting correlated D
meson pairs. Solid experimental measurements in pp and light-ion
collisions, at the same energy, are crucial for detailed studies of
changes in these azimuthal correlations, and should be performed as a
function of \pt. Upgrades of the STAR and PHENIX detectors with
micro-vertex capabilities~\cite{star_hft} and direct open charm
reconstruction should make these measurements possible.

Note that more than one pair of \DDbar\ could be produced in high-energy nuclear collisions.  At mid-rapidity, the total number of charm pairs is in the order of 2-4 and 6-10 at RHIC and LHC, respectively. The multiple production of un-correlated ${\rm D}$ mesons will give an additional background to the measured angular correlations. Our tests show that the effect can be removed by the mixed event method. The mixed-event subtracted correlation function can be compared with the calculated results. In addition, the next to leading order contributions, such as gluon splitting, to charm production may become more important in higher energy collisions. These \DDbar\ pairs are produced without a clear back-to-back correlations and will lead to a much weaker peak at 180 degrees. In order to understand the thermalization at RHIC, more theoretical analyses in this direction are called for.

{\it Acknowledgements}
After finishing this work, P.-B.~Gossiaux informed us about his
similar unpublished studies. We thank C.~Louren\c{c}o and H.K. W\"ohri for valuable comments and several suggestions.
This work has been supported by GSI and BMBF, in part by the U.S.
Department of Energy under Contract No. DE-AC03-76SF00098 and the NSFC Grants No. 10428510.


\vfill\eject
\end{document}